\documentclass[aps,prb,twocolumn,groupedaddress]{revtex4}                            
\usepackage{graphicx} 
\usepackage{epsfig} 
\usepackage{float}
\usepackage{amsmath}
\begin{document}

\title{Magnetic Coupling Between Non-Magnetic Ions: Eu$^{3+}$ in EuN
and EuP} \author{M. D. Johannes$^{\dag}$ and W. E. Pickett$^{\ddag}$}
\affiliation{$^{\dag}$Code 6391, Naval Research Laboratory,
                Washington, D.C. 20375}
\affiliation{$^{\ddag}$Department of Physics, University of California Davis,
                Davis CA 95616}

\begin{abstract} We consider the electronic structure of, and magnetic
exchange (spin) interactions between, nominally nonmagnetic Eu$^{3+}$
ions ($4f^6$, S=3, L=3, J=0) within the context of the rocksalt
structure compounds EuN and EuP.  Both compounds are ionic [Eu$^{3+}$;
$N^{3-}$ and $P^{3-}$] semimetals similar to isovalent GdN.  Treating
the spin polarization within the $4f$ shell, and then averaging
consistent with the J=0 configuration, we estimate semimetallic band
overlaps (Eu $5d$ with pnictide $2p$ or $3p$) of $\sim 0.1$ eV (EuN)
and $\sim 1.0$ eV (EuP) that increase (become more metallic) with
pressure. The calculated bulk modulus is 130 (86) GPa for EuN (EuP).
Exchange (spin-spin) coupling calculated from correlated band theory
is small and ferromagnetic in sign for EuN, increasing in magnitude
with pressure. Conversely, the exchange coupling is antiferromagnetic
in sign for EuP and is larger in magnitude, but decreases with
compression.  Study of a two-site model with $\vec{S_1} \cdot
\vec{S_2}$ coupling within the $J=0, 1$ spaces of each ion illustrates
the dependence of the magnetic correlation functions on the model
parameters, and indicates that the spin coupling is sufficient to
alter the Van Vleck susceptibility.  We outline a scenario of a
spin-correlation transition in a lattice of $S=3, L=3, J=0$
nonmagnetic ions. \end{abstract} \maketitle \today

\section{Introduction} The readily observable behavior of the angular
momentum and associated magnetic moment of rare earth ions is one of
the more obvious successes of the quantum mechanical description of
atomic magnetism.  Hund's rules for the total angular momentum $\vec J
= \vec L + \vec S$ in terms of orbital filling (giving orbital $L$ and
spin $S$ angular momenta) are simple and successful. The most striking
prediction is when the angular momenta in a spin polarized $S\ne 0$
and orbitally polarized $L\ne 0$ ion conspire to give a nonmagnetic
ground state $J = 0$.  The canonical example is Eu$^{3+} f^6$: $S=3,
L=3, J=0$. Whereas Eu$^{3+}$ is nonmagnetic experimentally
(unresponsive to magnetic fields, treatment of the high spin
polarization $\vec S^2 = 12$ (we use $\hbar$=1) is essential for
obtaining any reasonable model of the electronic structure.

The orbital polarization and associated moment $\vec m_L = \mu_B \vec
L$ is a secondary effect as far as the determination of the electronic
structure is concerned, because the magnetic coupling proceeds via the
effect of the spin moment $\vec m_S = 2\mu_B \vec S$ on the electronic
structure and on neighboring spins (via exchange coupling).  $\vec L$
is however central in determining $\vec J$ and the total moment, which
for Eu$^{3+}$ is large: $|\vec M| = \mu_B |(\vec L + 2 \vec S)| =
\mu_B |(\vec J + \vec S)| = \mu_B |\vec S| \approx 3.5 \mu_B$
($\langle\vec{M}^2 \rangle_{J=0}=12\mu_B^2$).  Of course, in the $J$=0
ground state the moment has zero projection along any axis including
any applied field, hence the ``nonmagnetic'' character of the ion.
These considerations do not change the fact that the ion has a large
intrinsic spin polarization, and that there will be exchange
(spin-spin) coupling between neighboring spins; $\langle\vec
S_i\rangle$=0 does not imply $\langle \vec S_i \cdot \vec
S_j\rangle$=0.  Such coupling may be difficult to observe because the
most direct means of observing magnetic coupling is via response to a
magnetic field.

In this paper we address some of the questions raised by the
interaction between these strongly polarized yet nonmagnetic ions
within the context of two of the simplest solid state realizations,
the binary compounds EuN and EuP. EuN is a semimetal with a simple
rocksalt structure.  Although it has been known for fifty years, there
is little characterization of this compound in the literature.  
Reported values of the lattice constant are consistent at
5.020$\pm$0.006~\AA.\cite{klemm,JKKAG73,brown,jacobs}.  Only a very
small non-stoichiometry range exists, although magnetic Eu$^{2+}$
impurities can occur in sufficient concentration in some samples to
mask the susceptibility of the Eu${3+}$ ion in EuN.\cite{didchenko}
There is also sparse literature for EuP, a compound both isostructural
and isovalent to EuN. The lattice constant of EuP is measured
\cite{gordienko} to be 5.756 $\AA$. The 15\% increase in lattice
constant (50\% larger volume) compared to EuN has a significant effect
on the exchange interaction, probably by shifting the balance between
competing mechanisms of exchange coupling.

Although any projection $\langle\vec S_i\rangle$ vanishes, when two
neighboring ions are coupled the spin-spin {\it correlation} $\langle
\vec S_1 \cdot \vec S_2 \rangle$ should be nonzero.  Correspondingly
there will be a term in the microscopic Hamiltonian $K \sum_{<ij>}\vec
S_i \cdot \vec S_j$ for all interacting pairs.  (We denote the
exchange coupling by $K$ and retain $J$ for other uses.)  This
coupling in itself breaks the spherical symmetry of the ion, coupling
the $J$ = 0, 1, 2,.., L+S states and allowing magnetic behavior to
emerge.  This effect is already seen in the magnetic susceptibility of
the Eu$^{3+}$ ion, where the magnetic field breaks the spherical
symmetry, mixes $J$=1 with $J$=0 character, and gives rise to the van
Vleck susceptibility \cite{TH98,CKJ85,BAMM97,SKPMR70}. One may
consider then whether this exchange coupling results in a phase
transition; that is, a coupling strength or temperature for which the
spin correlation length diverges.  If the $J$ = 0 state of each ion is
predominant, there will be only a weak magnetic signal and also minor
entropy [$k_B$log(2$J$+1)$\rightarrow$0]; this is the observed
nonmagnetic behavior of the Eu$^{3+}$ solid.  Vanishing entropy
implies that ordering could not be seen thermodynamically; on the
other hand, with free energy devoid of magnetic entropy ordering could
occur at much higher temperature.  Neutron scattering, which has
sensitivity to the difference between spin and orbital moments, should
be able to detect long range anti-aligned ordering of the spins (hence
of the magnetic moments) because of the AFM Bragg peak.  This still
primarily $J$=0 lattice of ions may therefore show a new kind of phase
transition in which the $\langle\vec S_i \cdot \vec S_j\rangle$
correlation length diverges but with small (perhaps negligible)
decrease in entropy. As the $J$=1 state gets mixed in, by exchange
coupling and by crystalline anisotropy, magnetic behavior will assert
itself and the transition would become easier to detect.  This process
can be controlled, and the tendencies enhanced, by pressure-induced
volume reduction.
 
In this paper we begin to pursue this line of reasoning
quantitatively.  We first apply density functional based correlated
band theory (LDA+U) to study the electronic structures of EuN and EuP
and their equations of state.  In the process we investigate the
extent to which correlated LDA+U results can reproduce Hund's rules
(more specifically, the $z$-components of the angular momenta). We
then apply conventional procedures to express the coupling in terms of
a Heisenberg exchange coupling constant $K$ between nearest neighbors.  
Using this exchange coupling, the known spin-orbit coupling (SOC)
constant, and a single-ion anisotropy parameter, we look at a toy
model of coupled ions, each of which has the $S=3, L=3, J=0$ ground
state of Eu$^{3+}$.  Specifically, we consider whether spin-spin
correlations between these J=0 ions can be detected, and whether they
may in fact lead to a phase transition (divergent correlation length).

\begin{figure}[tb] \begin{center} \includegraphics[width =
.95\linewidth]{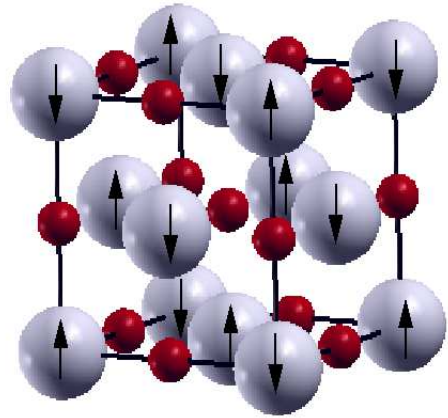} \caption{The AFMII spin ordering of a
rocksalt structure compound, such as is common in the transition metal
monoxides.  All second nearest neighbor spins, which are coupled by
the 180$^{\circ}$ superexchange coupling, are anti-aligned in this
ordered phase.} \end{center} \end{figure}

\section{Calculational Methods}

For Eu as for most rare-earth ions it is necessary to use the LDA+U
method (or another orbital-functional approach) to separate the
occupied $4f$ orbitals from the unoccupied ones, because LDA
invariably results incorrectly in narrow partially filled $4f$ bands
straddling the Fermi level. For our LDA+U calculations, we have chosen
a value $U$= 8 eV for the effective Coulomb repulsion and a value of
$J^H$=1 eV for the intra-atomic 4$f$ exchange interaction on the Eu
ion.  These values were chosen by comparison with other Eu
compounds.\cite{jan1,jan2} Since U is so large, the specific value of
U should not substantially change the conclusions about magnetic
coupling which are obtained. The value of J$^H$ is typical of what is
used for rare earth ions, and the effects of changing it have been
found to be small. The non-intuitive effect on the electronic
structure of the ``exchange'' J$^H$ as it arises in the LDA+U method
will be discussed in a Sec. III C.

All band structure calculations were performed using the
program\cite{Wien2k} Wien2k, which employs an APW+lo (augmented plane
wave plus local orbitals) basis set.  This allows for a lower plane
wave cutoff than is necessary for LAPW calculations, and consequently
the RK$_{max}$ was set to 7.00. We used the LDA exchange-correlation
parameterization of Perdew and Wang \cite{JPP92} and added a Hubbard U
according to the prescription of Anisimov {\it et al} \cite{VIA91,
VIA93}.  In this scheme, the double counting terms are subtracted
assuming a fully-localized (i.e. atomic) limit. \cite{AGP+03} The APW
sphere values were set to 2.1 a.u.  for Eu and 1.8 a.u. for N, in EuN
and to 2.4 a.u. for Eu and 2.2 a.u. for P in EuP.  These choices
resulted in negligible core charge leakage in both cases.

Band structure methods are seriously limited in the treatment of rare
earth ions due to their classical treatment of the spin and orbital
angular momentum (and for other reasons).  As usual, we assign a
direction (local $z$-axis) to $\vec S, \vec L$ and deal explicitly
with only the corresponding projections $S_z, L_z$; one might say that
we go into the reference frame of the spin on one Eu ion.  The closest
approximation to the $J$=0 state is then $S_z=3, L_z=-3$, which
uniquely specifies the occupation of one-electron $4f$ orbitals.  One
challenge the LDA+U method faces is this:  Hund's rules specify (1)
coupling all spins to maximum S, (2) coupling all orbital moments to
maximum L, and only then (3) considering the effects of spin-orbit to
obtain J.  One electron approaches, on the other hand, typically
incorporate spin-orbit coupling at the single electron level $\vec s +
\vec \ell = \vec j$.

Crystal field effects in both EuN and EuP are likely overestimated by
the LDA, resulting in an orbital moment that is ``over-quenched".
Comparing energies of the $S_z=3, L_z=-3$ state and a mixed state with
a diminished orbital moment of L$_z$ = -1.5, we find an energy
difference of about 25 mRy for both EuN and for EuP.  The J$_z$=0
state {\it is} (meta)stable and at some volumes the system converges
preferentially to this state, despite its somewhat higher total
energy. Such energy differences in LDA+U sometimes seem to approximate
excitations of the $4f$ shell.\cite{shick1} For these Eu pnictides and
at the current level of understanding, these energy differences should
probably not be taken too seriously; furthermore, the only noticeable
changes in the band structure between the two orbital configurations
occur in the splittings between occupied $4f$ states (see Section III
C).  We therefore use the lower energy state in our interpretations,
keeping in mind that, in reality, the orbital moment is likely to be
larger than our calculated moment.

It is unclear from the outset whether the exchange coupling will favor
alignment (ferromagnetism FM) or anti-alignment, AFMII, for example.  
(We use FM and AFM to refer to the relative spin alignment in our
calculations.) While the great majority of magnetic insulators are
AFM, in particular the transition metal monoxides, the rocksalt
structure EuO and Eu chalcogenides are FM or become FM under
pressure.\cite{jan2} In the calculations we present, we double the
unit cell to investigate the energetics of AFMII ordering. To obtain
an AFM solution, AFM symmetry of the wavefunctions and of the LDA+U
occupation matrix has been imposed. Occasionally the AFM symmetry
constraint was released to ensure that the solution remains stable
without constraint.

\section{Electronic Structure}

\subsection{E\lowercase{u}N } 
\begin{figure}[tb] \begin{center} \includegraphics[width =
0.99\linewidth]{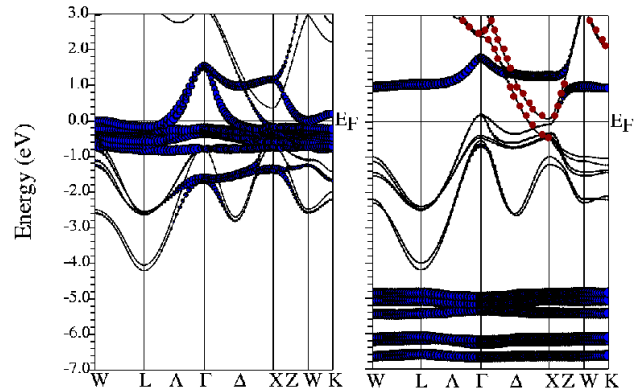} \caption{(color online) Band structures of
EuN in the simple unit cell with the $4f$ character highlighted by
broadened symbols. {\it Left panel:} the incorrect metallic solution
given by LDA.  {\it Right panel:} The LDA+U solution has six filled
majority $4f$ bands and a single empty $4f$ majority spin band,
separated roughly by the value of U.  The occupied $4f$ band
splittings (heavy lines) are discussed in the text. The Eu $5d$
conduction band dips below the Fermi energy to cross the valence N $p$
bands at the $X$ point; Eu $5d$ character is denoted by the (red)
circles.} \label{unou} \end{center} \end{figure}

As mentioned above, LDA is qualitatively
incorrect for the localized $4f$ states of EuN, and LDA+U must be used
to separate occupied from unoccupied $4f$ states.  In Fig. \ref{unou},
the LDA and LDA+U band structure results for the majority bands of FM
EuN are displayed.  The LDA results are qualitatively incorrect and
will not be discussed. The N $2p$ bands lie in the range -4 to 0 eV.  
The lower conduction bands are of Eu $5d$ character, disrupted by the
single empty majority spin $f$ band, and overlap the N $2p$ bands at
the $X$ point of the zone to form a semimetallic band structure. The
$f^6$ S=3 configuration is represented by six flat $4f$-bands in the
-4.5 eV to -7 eV range separated roughly by U from the single
unoccupied majority $4f$ state lying 1 eV above E$_F$.  The $4f$ bands
are so localized that they do not overlap even with neighboring $4f$
orbitals, making direct exchange negligible.  As we will show, the Eu
$5d$ states are considerably extended and provide one mechanism for
coupling of the localized spins (and hence the moments) on the Eu
ions.

The effects of the spin polarization of the localized $4f^6$ shell on
the electronic structure are considerable and nonintuitive.  Via the
on-site Hund's coupling (contained in the LDA exchange potential) the
Eu $5d$ bands are polarized with the same orientation, with an
exchange splitting of 0.6 eV.  The N $2p$ bands are also polarized,
but in the opposite direction.  As a result, the majority $2p$ bands
lie {\it above} the minority bands, and the induced magnetization on N
is negative.  While the majority Eu $5d$ band overlaps the majority
$2p$ band at E$_F$, there is a gap between the respective minority
bands.  Hence FM EuN is half metallic within conventional LDA+U band
theory.

\begin{figure} \includegraphics[width = 0.99\linewidth]{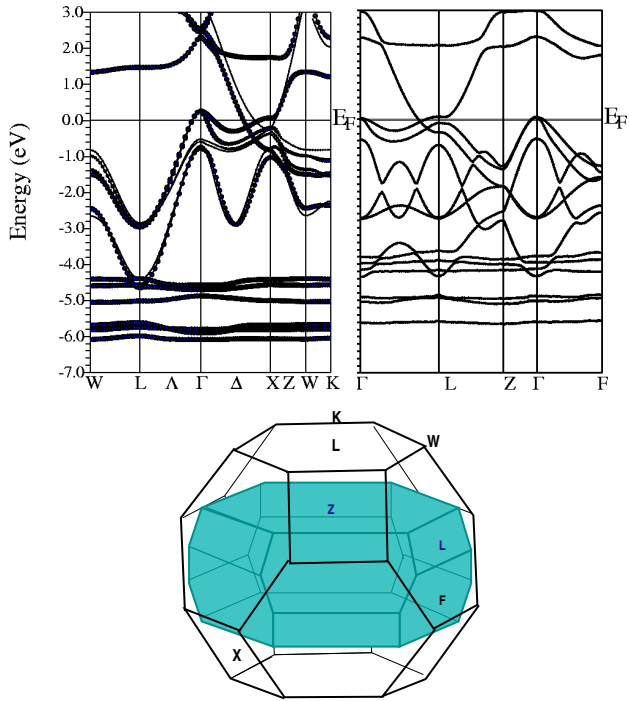}
\caption{A direct comparison of EuN FM and AFM band structures at a
4\% compressed lattice constant (15 GPa pressure). The FM
configuration (left) is shown in the FCC unit cell with all majority
spin states highlighted, while the AFM configuration (right) is shown
in doubled rhombohedral cell.  The rhombohedral (shaded) Brillouin
zone is shown inside the larger BZ of the FCC unit cell below. The
important Eu $d$ overlap with the N $p$ bands occurs at the $X$ point
in the large BZ, corresponding to the $L$ point of the small BZ.}
\label{c2r} \end{figure}

A comparison of the FM band structure of EuN in a compressed
face-centered cubic (FCC) unit cell and the AFM band structure in a
similarly compressed, doubled, rhombohedral unit cell is shown in Fig.
\ref{c2r}.  Compression moves the Eu $5d$ bands down relative to the N
$2p$ bands, increasing the filling of $5d$ states at the $X$ point
($L$ in the rhombohedral cell) and beginning to fill a minority Eu
$5d$ pocket.  Thus half metallicity is lost under moderate compression
(pressure).  The compensating holes go into the $2p$ band, at $\Gamma$
and at the $X$ point.  The remaining complete spin polarization of the
N $2p$ holes may be important for the $4f-4f$ exchange processes (Sec.
IV).

The only previous calculations on EuN were by Horne {\it et
al.}\cite{SIC} using the self-interaction-corrected LDA (SIC-LDA)
approach implemented within the linearized muffin-tin orbital band
structure code.  Horne {\it et al.} reported only the DOS for FM spin
alignment, with the results being significantly different from what we
have obtained, as the considerable differences in the methods might
suggest.  Regarding the Eu $5d$ and N $2p$ bands, they obtained a
semiconducting result with a separation of 1.4 eV, whereas LDA+U gives
an overlap of several tenths of an eV for FM alignment.  The N $2p$
bandwidth is 15\% narrower than the LDA+U value of 4.5 eV. IN SIC-LDA
the unoccupied majority $4f$ orbital lies at the very bottom this gap,
mixes strongly with the N $2p$ states, acquires a 1 eV width and
becomes slightly occupied.  The SIC-LDA result thus is a half metal
(for FM alignment).  The minority $4f$ bands (unoccupied) lie about 4
eV above E$_F$, quite similar to what we find from LDA+U.  The
majority $4f$ states in SIC-LDA lie about 12 eV below E$_F$ with 0.5
ev width, whereas those obtained from LDA+U are centered only 5.5 eV
below E$_F$ with a 2 eV width. As we discuss in Sec. III C, the $4f$
state splittings in LDA+U are a result of the anisotropy of the
exchange interaction. These various differences generally reflect the
main difference in the methods: SIC-LDA lowers (very strongly) the
occupied $4f$ states, whereas LDA+U lowers occupied and raises
unoccupied $4f$ states each by some fraction of U.

Other rare earth nitrides have been reported in experimental studies
to be semimetals \cite{ScN,GdN,YbN} as our results give for EuN, while
the Group IIIA mononitrides (BN, AlN, GaN, InN) are known to be wide
gap semiconductors. Calculations of the electronic structure of the
closely related compound GdN have been given by Petukhov {\it et
al.},\cite{petukhov} who used an atomic sphere approximation for the
potential and treated the $4f$ states corelike (thus removed from the
band structure problem).  These calculations gave a small (0.1 eV)
overlap between the valence band maximum at $\Gamma$ involving N $2p$
states, and a conduction band minimum at X comprised of strongly Eu
$5d$ character.  This band structure is very much like what we obtain
for EuN (as well as for antiferromagnetic EuP below).  By averaging
over the spin splitting (and also surveying the AFM bands) we deduce a
semimetallic band overlap of $\sim 0.1$ eV for EuN, and $\sim 1.0$ eV
for EuP, both of which increase under pressure.

\begin{figure}[tb] \begin{center} \includegraphics[width =
.99\linewidth]{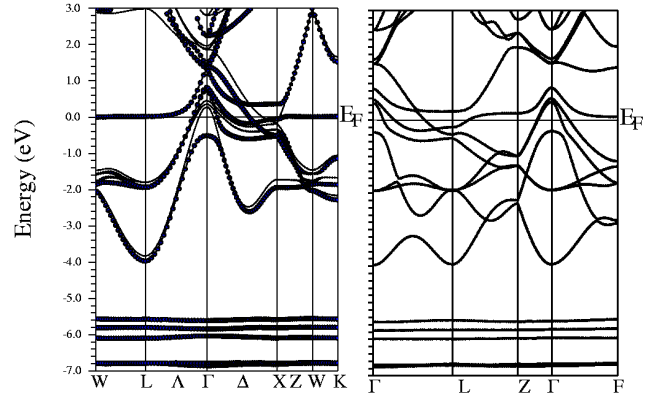} \caption{(color online)  {\it Left panel:}
bands of FM EuP in the FCC cell with majority spin character
highlighted. {\it Right panel:} AFM EuP in the rhombohedral cell at
equilibrium volume.  Note that the $4f$-$3p$ mixing near E$_F$ is so
strong in the AFM case that the unoccupied $4f$ band above E$_F$ is
not readily obvious.}
 \label{AFMbands}
\end{center}
\end{figure}

\begin{figure}[tbp] \begin{center} \includegraphics[width =
0.85\linewidth, angle = 270]{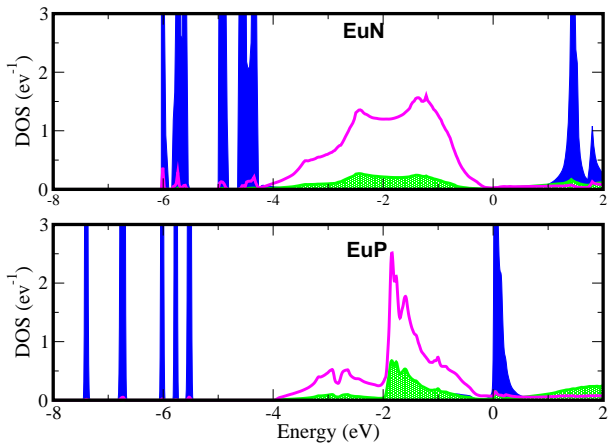} \caption{ (color online) A
partial DOS plot for FM EuN and AFM EuP. {\it Top panel:} The
orbitally resolved $5d, 4f$ states of Eu and $2p$ states of N at a
compression of 12\% by volume (15 GPa).  The unoccupied $4f$ band is
more than 1 $eV$ above the filled N $2p$ bands.  The bottom of the
valence $5d$ band is filled, providing the mechanism for
superexchange.  {\it Bottom panel:} At equilibrium volume, the
unoccupied $f$ band sits immediately above E$_F$, and the occupied
bands are completely separated from the filled P $3p$ bands.}
\label{dos} \end{center} \end{figure}

\subsection{E\lowercase{u}P} The band structures of FM- and
AFMII-ordered EuP are shown in Fig. \ref{AFMbands}.  The main features
of EuN remain in EuP, specifically the character of the valence and
conduction bands and the semimetallic band crossing at the $X$ ($L$)
point.  In EuP, the valence P $3p$ bands that determine E$_F$ are
higher in energy (relative to the $4f$ bands) than the analogous $2p$
bands in EuN.  Conversely, one may adopt the viewpoint that the Eu
$5d$ and $4f$ bands are shifted downward relative to the P $3p$ bands
(see Fig. \ref{dos}).  One feature clearly relevant to magnetic
interactions is $4f-3p$ hybridization.  The unoccupied majority $4f$
band sits just above $E_F$ and mixes strongly with the P $3p$ band
near $X$ along the $\Gamma-X$ line.  Holes in the $3p$ band are
consequently near the $\Gamma$ point only and {\it not} from the
majority spin band near $X$ which is pushed down by hybridization with
the $f$ band.  In contrast to EuN, EuP with FM alignment is not a
half-metal and significantly, the $3p$ holes are only weakly
spin-polarized.

\subsection{$4f$ Occupation and Hund's 2nd rule} It should be
understood at the outset that any band picture of the $4f$ states has
serious intrinsic limitations that have been recognized for some time
and occasionally addressed.\cite{ulf} A band structure is hoped to
represent, in a mean field manner, the electron removal spectrum for
occupied states and the electron addition spectrum for unoccupied
states.  The open $4f$ shell is a highly correlated local system:
individual spins are strongly coupled and aligned to form the total
spin $S$, orbital occupations are correlated to give $L$, and Hund's
third rule (SOC) gives the ground state $J$.  In such ions there is a
manifold of low energy excitations in which the particle number
remains unchanged ($J\neq 0$ multiplets at higher energy). These
multiplet excitations, often affected strongly by crystal fields, lie
outside the realm of band structures, even in a `zeroth order'
description.

Density functional theory (DFT), which gives rise in the process of
energy minimization to the Kohn-Sham band structure, is itself a many
body theory for the system energy.  There have been a few indications
that the LDA+U approach is useful in modeling (perhaps predicting)
some of these particle-number conserving
excitations.\cite{shick1,shick2} More specifically, one can ask
whether the LDA+U approach is capable of reproducing Hund's rules,
{\it i.e.} obtaining the lowest energy for the correct configuration
$L, S, J$. Solovyev {\it et al.} have addressed this
question,\cite{solovyev} and Gotsis and Mazin have obtained
encouraging results\cite{gotsis} for some Sm intermetallic compounds.

The separations of the filled flat $4f$ bands in Figs. \ref{unou},
\ref{c2r}, \ref{AFMbands} are due to the strong orbital-dependence of
the exchange interaction, whose full anisotropy is incorporated into
the LDA+U code.  These differing energy shifts are determined by the
Hund's exchange constant J$^H$. To clarify the origin of the splitting
between the f-bands, we present an explicit example of the potential
matrix elements used in the LDA+U procedure for the case where the
density matrix is diagonal, $n^{\sigma}_{mm'} = n^{\sigma}_m
\delta_{mm'}$ (which avoids uninstructive complications that arise in
the general case).  Then the only matrix elements which enter the
LDA+U energy are the direct and exchange integrals of the form
\begin{align} {\cal U}_{mm'} = \langle mm'|V_{ee}|mm' \rangle \\ {\cal
J}^H_{mm'} = \langle mm'|V_{ee}|m'm \rangle \notag \end{align} In the
atomic limit these are given simply in terms of the $\ell = 3$ Slater
integrals F$_0$, F$_2$, F$_4$, F$_6$ and angular factors.  In the
LDA+U method these are represented in terms of two constants:
``direct'' $U$ and ``exchange'' $J^H$ \begin{align} U = F^0; \qquad
J^H = (F_2 + F_4)/14 \\ \text{and the ratio} \qquad F_4/F_2 \sim 0.625
\qquad F_6/F_2 \sim 0.494 \notag \end{align} With U = 8 eV and J$^H$ =
1 eV as we have used, the corresponding Slater integrals are: F$_0$ =
8.0 eV, F$_2$ = 11.92 eV, F$_4$ = 7.96 eV, and F$_6$ = 5.89 eV.  
(F$_0$ is strongly screened in the solid so atomic values cannot be
used directly.)
                                                                                         
If J$^H$ is set to zero, these simplify to an isotropic repulsion
${\cal U}_{mm'}$ = F$_0$, and to a diagonal and orbital-independent
${\cal J^H}_{mm'}$ = F$_0$ $\delta_{mm'}$, the latter simply removing
the self-interaction.  Thus for J$^H$=0 no distinction between
different orbitals is made, and filled orbitals will all be shifted
downward equally, while unfilled orbitals will be shifted upward
equally. An important and unsatisfactory result is that the energy
differences based on orbital quantum number $m$ (which give rise to
Hund's 2nd rule splittings in the atomic limit) are absent when
J$^H$=0.

With J$^H \neq$ 0 the diagonal elements become orbital dependent, and
each occupied orbital obtains (self-consistently) its own downward
shift in energy.  To illustrate the strong orbital dependence
(anisotropy) we present explicitly the ${\cal U}_{mm'}$ and ${\cal
J}_{mm'}$ matrices for Eu in Appendix I. The resulting differences in
eigenenergies can be pronounced, as in Fig. \ref{AFMbands} where
different occupied $4f$ bands are separated by as much as 2 eV. J$^H$
therefore has two roles: (1) it provides the anisotropy of the direct
Coulomb repulsion ${\cal U}_{mm'}$ (which is independent of $m,m'$ if
$J^H$=0), and (2) it has the more commonly understood aspect of
providing the intra-atomic exchange coupling ${\cal J}_{mm'}$, $m\neq
m'$.

\subsection{Equation of State} 

We fit the energy vs. volume curves for the FM/AFM spin configuration
of EuN/EuP to an equation of state \cite{EOS}: \begin{equation} E = a
+ bV^{-1/3} + cV^{-2/3} +cV^{-1} \end{equation} and extracted the LDA
equilibrium constants and bulk moduli.  EuN has a calculated
equilibrium volume of 406.9 $a.u.^3$, 5\% smaller than experiment
(consistent with typical LDA error) and a bulk modulus, B=130 GPa.  
The volume of EuP is 612.6 $a.u.^3$, consistent with the larger
P$^{3-}$ ion, and again is 5\% smaller than experiment.  It has a bulk
modulus of 86 GPa, making it much softer than EuN. The lattice
constants of EuN and EuP have been calculated previously by Horne {\it
et al.},\cite{SIC} whose SIC results are very similar to ours.

\section{Exchange Energy and Interactions} In the rocksalt structure
each Eu ion has potential exchange coupling to its twelve nearest Eu
neighbors through an intermediate pnictide ion at a 90$^{\circ}$
angle, and to each of its six second neighbors at a 180$^{\circ}$
angle.  The Goodenough-Kanamori superexchange rules lead one to
anticipate a ferromagnetic nearest neighbor interaction, while the
second nearest neighbor interaction will be antiferromagnetic.  These
guidelines are based on perturbation expansions appropriate for
insulators, and may be misleading here where other mechanisms also
arise (RKKY interactions through the electron and hole carriers,
specifically).  We probe some energetics of the superexchange
processes by calculating the energy differences between ferromagnetic
and AFMII spin alignments, using the same rhombohedral cell to improve
numerical precision.  For the experimental lattice constant of EuN,
there is a small energy difference $\Delta E$ = E$_{FM}$ - E$_{AFM}$ =
-4.9 meV.  $\Delta E$ increases in magnitude with compression and
reaches a value of -10.1 meV when the lattice parameter is reduced by
$\approx$ 4\%.  In EuP, the energy differences are somewhat larger and
of the opposite sign, with $\Delta E$= 12.8 meV at equilibrium volume
decreasing to $\Delta E$ = 9.2 meV at a compression by 4\% of the
lattice constant. The calculated ground state of EuN is therefore FM,
while that of EuP is antiferromagnetic.  Our single total energy
difference is only sufficient to give a single exchange constant. As
can be seen in the equation below, the effective exchange constant
(which includes {\it all} exchange interactions between {\it all}
neighbors) may be attributed solely to the first neighbors or solely
to the second neighbors, resulting in an identical expression. Using
nearest neighbors only in a classical S=3 Heisenberg model gives for
the two magnetic states: \begin{equation} E_{AFM} = 0, \qquad E_{FM}
=6 KS^2, \qquad \Delta E = 6KS^2. \end{equation} Using second nearest
neighbors only gives:  \begin{equation} E_{AFM} = -3KS^2, \qquad
E_{FM} =3 KS^2, \qquad \Delta E = 6KS^2. \notag \end{equation}

\begin{figure} \includegraphics[width =
0.80\linewidth,angle=270]{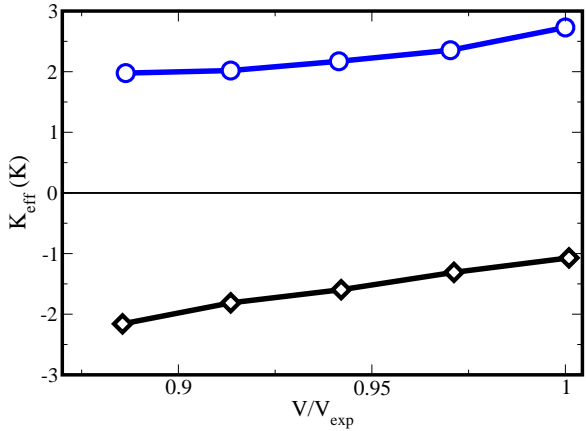} \caption{The inferred exchange
constant, K, for EuN and EuP. {\it top panel:} K for EuP is positive
(antiferromagnetic order) and decreases with pressure. In analogy with
the $3d$ transition metal monoxides, AFMII order may more likely be
due to next-neighbor AFM coupling than to near-neighbor AFM coupling.  
{\it lower panel:} K for EuN is negative (ferromagnetic order) and
increases in magnitude with pressure.} \label{KK} \end{figure}

Extracting the exchange constant from these formulas gives K$_{EuN}$ =
-0.09 meV (1K) at zero pressure and K$_{EuN}$ = -0.19 meV (2.2K) at
4\% reduced lattice constant or 15 GPa .  For EuP, the exchange
interaction is somewhat larger, K$_{EuP}$ = 0.24 meV (2.8 K) at
equilibrium pressure, and opposite in sign.  It also has the opposite
pressure dependency which, due to the smaller bulk modulus, produces a
decrease of nearly 25\% in K$_{EuP}$ at a pressure of 7.5 GPa (see
Fig.  \ref{KK}).  We observe that for the larger compound, EuP, the
effective interaction between ions decreases with increasing pressure
despite the decreased distance between neighbors, possibly due to
decreasing competition from first neighbor exchange.  EuN is firmly
within the FM regime and pressure has the expected effect of
increasing the coupling.  Recent pressure studies show that EuN
retains the rocksalt structure up to 70 GPa. \cite{BM} Assuming that
the lattice constant and $K$ maintain a linear relationship at high
pressure (Fig. \ref{KK} indicates that in reality it is slightly
sub-linear), we can extrapolate that $K$ = 0.3 meV (3.5 K) at 70 GPa.  
As we will show in the next section, an exchange coupling of this
magnitude can have observable effects on the susceptibility. Despite a
difference in Eu valency, these results are quite consistent with the
trends found for the series of Eu$^{2+}$X compounds \cite{jan2},
namely that compounds with a large lattice constant are AFM, while
compounds under high pressure or with smaller lattice constants are
FM.  The exchange constants of Ref. \onlinecite{jan2} were resolved
into $K_1$ and $K_2$ separately, both of which are comparable in
magnitude to our single effective $K$.

\section{A Toy Model Study} Since the states of ions are most readily
described in terms of their total angular momentum $\vec{J}_i$ it is
natural and customary to use the effective Hamiltonian
$K_{eff}\vec{J}_1 \cdot \vec{J}_2$ for the coupling.  However, the
actual exchange interaction proceeds via the spins $\vec{S}_1$ and
$\vec{S}_2$, so we begin this investigation by considering the more
microscopic model Hamiltonian expressed in terms of $\vec{S}_i$ and
$\vec{L}_i$.  Our two identical site Hamiltonian ($i$=1, 2) is
\begin{equation} H = K \mathbf{S}_1 \cdot \mathbf{S}_2 +
       \lambda \sum_i \mathbf{L}_i \cdot \mathbf{S}_i -
         \sum_i \mathbf{M}_i  \cdot \mathbf{B}  + 
       \sum_i (\mathbf{D} \cdot \mathbf{L}_i)^2.
\label{model}
\end{equation}

This Hamiltonian includes the interactions we expect to be important
in the investigation of magnetic behavior in EuN or EuP (although
interactions may extend to more neighbors).  We choose the spin-orbit
constant to be $\lambda$=323 cm$^{-1}$ = 40 meV, a typical value for
the $^7F_0 \rightarrow$$ ^7F_1$ transition observed by spectroscopic
\cite{JRGT+99,CGKB93,XYC,CKJ85} measurements on several Eu$^{3+}$
compounds.  The exchange parameter K will be varied to examine the
macroscopic effects of the exchange coupling, with the physically
relevant values taken from our LDA+U calculatios.  The most
conservative estimate is obtained by directly importing the $K$
derived by mapping energy differences onto the Heisenberg model.  
This gives an upper bound of $K/\lambda \sim$ 0.01 for EuN (under high
pressure) and for EuP at equilibrium volume.  It may be more realistic
to establish $K$ by directly forcing the toy model to reproduce the
density functional AFM-FM energy differences (per Eu ion pair).  This
gives a much higher value ratio of $K/\lambda $: 0.21 for EuP and 0.08
for EuN, both at their respective equilibrium volumes. The single-ion
anisotropy $\vec D$ is taken to be of similar magnitude (2 meV), with
the easy axis defining the $\hat z$ axis.

Our basis includes J=0 and three J=1 states on each of the two sites.  
The next J multiplet, $^7F_2$, is $2\lambda \sim 900 K$ higher in
energy than $^7F_1$, so it will not contribute to the low energy
regime we explore.  This basis of four states per site leads to a
$16\times 16$ Hamiltonian matrix. Matrix elements are readily
evaluated in the $|L,L_z;S,S_z \rangle $ basis;  the expansion of the
states $|J,J_z \rangle$ states ($\vec J = \vec J_1 + \vec J_2$) in the
$|L,L_z;S,S_z \rangle$ states is given in Appendix II.

For K=0, D=0, Hund's rules give the $J_i$=0 ground state on each site
(and trivially J$_{tot}$=0).  As discussed in the Introduction, any
projection of each magnetic moment $\vec{M_i}= \mu_B(\vec{L_i} +
2\vec{S_i})$, and therefore the total $\vec M = \vec M_1 + \vec M_2$,
vanishes for uncoupled spherical ions although its mean square value
is large. Exchange coupling K introduces mixing between between the
states of the two sites, but preserves the spin rotational symmetry of
the Hamiltonian.  Expectation values of both individual and total
magnetic moments (and individual angular momenta) therefore are zero,
but correlation functions such as $\langle \vec{S}_1 \cdot \vec{S}_2
\rangle$ and $\langle \vec{M}_1 \cdot \vec{M}_2 \rangle$ grow as the
strength of the coupling grows.  Single-site anisotropy $\vec D \neq
0$ or field $\vec B$ destroys the rotational symmetry of H and
introduces additional mixing of states.

\begin{figure}[tb] \begin{center} \includegraphics[width = 3.0 in,
angle=270]{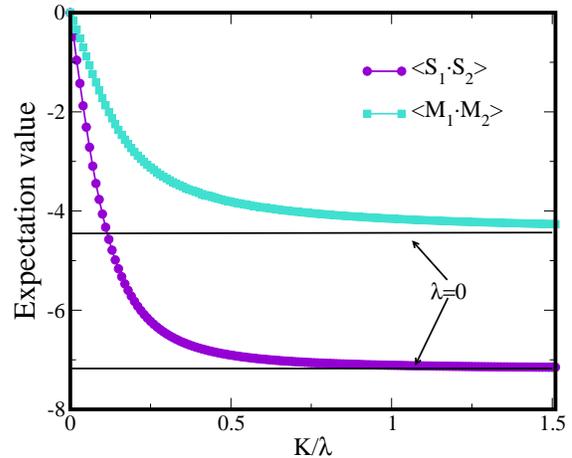} \caption{Alignment of spins and magnetic moment as
a function of $K/\lambda$.  The horizontal line in each case marks the
value of maximum possible alignment possible in the restricted Hilbert
space of the model} \label{corr} \end{center} \end{figure}

Fig. \ref{corr} shows the increase of the spin-spin and magnetic
moment correlation functions in the ground state as K increases.  In
the limit $K/\lambda \rightarrow \infty$, the model reduces to a two
site Heisenberg model, with a well known analytic solution: $\vec{S}_1
\cdot \vec{S}_2 = (S^2 - S_1^2 - S_2^2)/2 = -12 $, where $S=0$ in the
ground state and $S_i$=$L_i$=3. The restricted basis set we use does
not span the entire Hilbert space of the Heisenberg Hamiltonian and
the spin-spin correlation function therefore asymptotes to a smaller
value of -7.18.  The higher J multiplets necessary for $ \langle
\vec{S}_1 \cdot \vec{S}_2 \rangle $ to attain the analytical ground
state value are moved too high in energy by the scale of the
spin-orbit parameter, therefore -7.18 should be considered the maximum
anti-alignment of spins within this system. The asymptotic and
analytically calculated expressions for $ \langle {\bf M}_1 \cdot {\bf
M}_2 \rangle $ are also different because of the restricted basis set.

For small $K$ the order parameter grows linearly with K with steep
slope, and at $K/\lambda$ = 0.175, $\langle \vec S_1 \cdot \vec
S_2\rangle$ reaches half its maximum value, as shown in Fig.
\ref{corr}.  A similar conclusion holds for the moment-moment
correlation function. Rotational symmetry of the system is broken,
allowing a non-zero moment and ready detection, by single-ion
anisotropy $\vec{D} \neq 0$ or by application of a magnetic field
$\vec B$.

\begin{figure} \includegraphics[width = 3.2 in, angle=270]{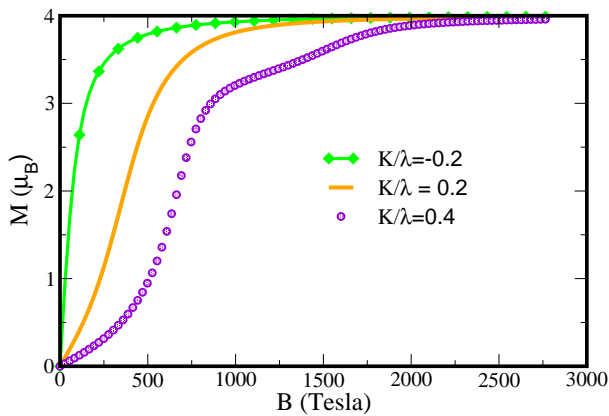}
\caption{Magnetic moment vs. applied magnetic field for a
thermodynamic ensemble of two site ``molecules''described by Eq.
\ref{model}.  The curves are sensitive to the value of the exchange
constant K, suggesting that this quantity can be probed with
experimental measurements of M vs. B.} \label{MB} \end{figure}

A calculation of $ \langle M_{\parallel} \rangle $ vs. $B$ in thermal
equilibrium, presented in Fig. \ref{MB}, shows two clear crossovers at
fields of $\sim$ 720 and 1450 Tesla.  These transitions exist only for
relatively high values of K and low temperatures and correspond to
qualitative changes in eigenvectors as the external B-field overcomes
the exchange and spin-orbit energies respectively.  At higher
temperatures or lower K/$\lambda$, every state in the Hilbert space is
present even in the ground state and the curve evolves smoothly.
Because fields of this magnitude are experimentally unattainable and
would be strong enough to mix in even higher J multiplets (J=2,3...),
only the low field part of the curve in Fig. \ref{MB} can be taken
seriously.  In low fields, $\langle M \rangle$ is linear in B with a
slope (susceptibility $\chi$) that is strongly dependent on the
exchange coupling.  This sensitivity suggests that the degree of
spin-spin correlation in a system containing Eu$^{3+}$ ions can be
probed by examining the temperature dependence of the magnetic
susceptibility.

\begin{figure} \includegraphics[width = 2.7 in, angle=270]{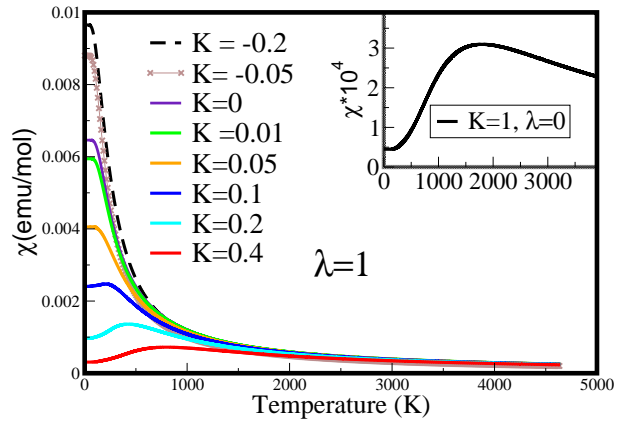}
\caption{The magnetic susceptibility of the two-site model with
various levels of spin-spin interaction.  The curve resembles a Van
Vleck paramagnetic susceptibility for small K, and smoothly evolves to
a shape of susceptibility typical of local AFM correlations when K
dominates.} \label{susc} \end{figure}

Fig. \ref{susc} shows the calculated $\chi(T;K,\lambda)$ curves for
different values of $K$ within our two site model.  For small $K$, the
curves show typical Van Vleck paramagnetic behavior \cite{vleck},
characterized by an initial flat region which quickly decreases
towards zero after a certain onset temperature.  The magnitude of the
susceptibility in the constant region of the curve and the onset
temperature of the decrease are both extremely sensitive to the
$K/\lambda$ ratio.  Susceptibility curves reported in Refs.
\onlinecite{TH98,CKJ85,BAMM97,SKPMR70} for Eu$^{3+}$ compounds are
very similar to those calculated within our model for small $K$.  For
larger values of $K$, beginning with $K/\lambda \sim$ 0.1, the curve
begins to develop a maximum at non-zero temperature which then
evolves, with growing $K$, more toward a standard AFM susceptibility
as shown in the inset of Fig. \ref{susc} for the limit $K/\lambda
\rightarrow \infty$. Since the $K/\lambda$ ratio is manipulable by
pressure, the effect of increasing $K$ (signalling increasing
spin-spin correlation) should be observable in the susceptibility
curve, even for the conservative estimates of $K/\lambda$ mentioned
earlier.

The first principles calculation indicates that a non-zero spin-spin
interaction of conventional strength does indeed exist.  Though the
energy difference between spin orientations for EuN is small at
ambient volume, its magnitude can be increased with pressure.  For
EuP, the energy difference is largest at equilibrium volume and is
sensitive to manipulation by pressure.  The model calculation shows
that any non-zero interaction results in nonvanishing spin correlation
$\langle \vec S_i \cdot \vec S_j\rangle$.

\section{Discussion} We have initiated the study of the electronic
structure of EuN and EuP specifically, and of the spin coupling and
resulting behavior of the $4f^6$ ion more generally.  The strong spin
polarization of the $4f$ shell, and the coupling of the moments
L+S=J=0 to produce a nonmagnetic ion, makes a prediction of the
physical electronic structure somewhat uncertain.  We obtain, however,
that both EuN and EuP are semimetallic due to overlap of the valence
pnictide $p$ bands with the conduction Eu $5d$ bands.  Recent
measurement of the resistivity of pressed powders of EuN show a
positive temperature coefficient, {\it i.e.} metallic or semimetallic
conduction.\cite{cornell}

We point out that even though the expectation value of any (vector)
spin may be zero, exchange coupling will give rise to nonvanishing
spin-spin correlations between coupled ions.  This coupling may lead
to an unconventional phase transition in which the spin correlation
length diverges but there is little or no signal in the magnetic
properties.  The determination of the ordering temperature acquires
new aspects: since entropy is (nearly) absent in both ordered and
disordered states, a different mechanism from simple energy-entropy
balance may control the ordering temperature.  One possibility is that
spin-phonon interactions promote disorder: raising the temperature
increases thermal vibrations, which modulates the interatomic
separations and hence modulates the exchange couplings dynamically.  
The phase transition would occur at a critical value of the exchange
coupling ``disorder.''

There are other Eu rocksalt compounds (divalent, $4f^7$) with FM
ordering, as well as those with AFM ordering.  In some cases, such as
EuSe, a transition between the two magnetic alignments can be driven
by manipulating the U parameter \cite{jan2} or by applying pressure.  
Based on our LDA+U results, a state in which the spin moments of Eu
ions are antiferromagnetically correlated is unlikely for EuN but
probable for EuP. For EuN, a hidden ferromagnetic correlation, or
ordering, of spins (accompanied by an identical, but oppositely
pointed arrangement of orbital moments) is energetically favored.  It
could be an experimental challenge to see the hidden ``ferromagnetic
order,'' which does not break any crystal symmetry.

\section{Acknowledgments} The authors are indebted to A. B. Shick for
many discussions on this work and for the code to produce the data
presented in Appendix I. We acknowledge discussions with Z. Fisk, J.
Lynn, S. M. Kauzlarich, J. Kune\v{s}, I. I. Mazin, D.A.
Papaconstantopoulos, P. Novak, and D. J. Singh. W. E. P. was supported
by DOE grant DE-FG03-01ER45876, DOE's Computational Materials Science
Network, and the SSAAP program at LLNL under grant DE-FG03-03NA00071.

\section{Appendix I}
With U = 8 eV and J$^H$ = 1 eV as used in our LDA+U calculations, the
direct and exchange matrix elements for diagonal occupation numbers
are given by the following tables  
(the matrix elements 
are ordered from $\ell_z$ = -3 $\rightarrow \ell_z$ = 3 from
left to right and up to down): 
\begin{align*}
{\cal U}_{mm'}=
\begin{bmatrix}
&&&&&& \\
9.4 & 7.8 & 7.2 & 7.1 & 7.2 & 7.8 & 9.4 \\
7.8 & 8.4 & 7.9 & 7.8 & 7.9 & 8.4 & 7.8 \\
7.2 & 7.9 & 8.7 & 8.4 & 8.7 & 7.9 & 7.2 \\
7.1 & 7.8 & 8.4 & 9.4 & 8.4 & 7.8 & 7.1 \\
7.2 & 7.9 & 8.7 & 8.4 & 8.7 & 7.9 & 7.2 \\
7.8 & 8.4 & 7.9 & 7.8 & 7.9 & 8.4 & 7.8 \\
9.4 & 7.8& 7.2 & 7.1 & 7.2 & 7.8 & 9.4 \\ \notag
\end{bmatrix}
\end{align*}

\begin{align*}
{\cal J}^H_{mm'} = 
\begin{bmatrix}
&&&&&& \\
9.4 & 1.55 & 0.95 & 0.53 & 0.48 & 0.37 & 0.74 \\
1.55 & 8.4 & 1.11 & 1.26 & 0.41 & 0.92 & 0.37 \\
0.95 & 1.11 & 8.7 & 0.50 & 1.90 & 0.41 & 0.48 \\
0.53 & 1.26 & 0.50 & 9.4 & 0.50 & 1.26 & 0.53 \\
0.48 & 0.41 & 1.90 & 0.50 & 8.7 & 1.11 & 0.95 \\
0.37 & 0.92 & 0.41 & 1.26 & 1.11 & 8.4 & 1.56 \\
0.74 & 0.37 & 0.48 & 0.53 & 0.95 & 1.55 & 9.4 \\ \notag
\end{bmatrix}
\end{align*}

\section{Appendix II}
We give here the parentage of the $|J,J_z)$ states in terms of the the product
states $|S,S_z;L,L_z> \equiv |S_z,L_z\rangle$ ($S=3, L=3$ are fixed.  For our
considerations only the $J=0$ and $J=1$ states are needed.
\begin{eqnarray*}
|0,0)&=&\frac{1}{\sqrt{7}}\Bigl[(-3,3\rangle -|-2,2\rangle +|-1,1\rangle 
-|0,0\rangle \\ \nonumber
&&+|1,-1\rangle -|2,-2\rangle +|3,-3\rangle \Bigr] \\ \nonumber
|1,-1)&=&\frac{1}{2\sqrt{7}}\Bigl[\sqrt{3} |2,-3\rangle -\sqrt{5}|1,-2\rangle 
+\sqrt{6}|0,-1\rangle \\ \nonumber
&&-\sqrt{6}|-1,0\rangle +\sqrt{5}|-2,1\rangle - \sqrt{3}|-3,2\rangle \Bigr] \\ \nonumber
|1,0)&=&\frac{1}{2 \sqrt{7}} \Bigl[-3|-3,3\rangle +2|-2,2\rangle -|-1,1\rangle \\ \nonumber
&&+|1,-1\rangle -2|2,-2\rangle +3|3,-3\rangle \Bigr] \\ \nonumber
|1,1)&=&\frac{1}{2\sqrt{7}} \Bigl[\sqrt{3}|3,-2\rangle -\sqrt{5}|2,-1\rangle 
+\sqrt{6}|1,0\rangle \\ \nonumber
&&-\sqrt{6}|0,1\rangle +\sqrt{5}|-1,2\rangle -\sqrt{3}|-2,3\rangle \Bigr] \\ \nonumber
\end{eqnarray*}
\bibliography{eunp2.bib}
\end{document}